\documentclass[onecolumn]{svjour2}                    % onecolumn
\smartqed  % flush right qed marks, e.g. at end of proof
\usepackage{graphicx}% Include figure files
\usepackage{dcolumn}% Align table columns on decimal point
\usepackage{bm}% bold math
\usepackage{amssymb}%
\usepackage{mathptmx}      % use Times fonts if available on your TeX system

\journalname{Foundations of Physics}
\begin{document}

\title{How Classical Particles Emerge From the Quantum World}
\titlerunning{How Classical Particles Emerge}

\author{Dennis Dieks and Andrea Lubberdink}

\institute{D. Dieks \at
              Institute for the History and
Foundations of Science, Utrecht University\\ P.O.Box 80.010, 3508
TA Utrecht, The Netherlands \\
              \email{d.dieks@uu.nl} }

\date{Received: date / Accepted: date}

\maketitle

\begin{abstract}
The symmetrization postulates of quantum mechanics (symmetry for
bosons, antisymmetry for fermions) are usually taken to entail
that \emph{quantum particles} of the same kind (e.g., electrons)
are all in exactly the same state and therefore indistinguishable
in the strongest possible sense. These symmetrization postulates
possess a general validity that survives the classical limit, and
the conclusion seems therefore unavoidable that even classical
particles of the same kind must all be in the same state---in
clear conflict with what we know about classical particles. In
this article we analyze the origin of this paradox. We shall argue
that in the classical limit classical particles \emph{emerge}, as
new entities that do not correspond to the ``particle indices''
defined in quantum mechanics. Put differently, we show that the
quantum mechanical symmetrization postulates do not pertain to
\emph{particles}, as we know them from classical physics, but
rather to indices that have a merely formal significance. This
conclusion raises the question of whether the discussions about
the status of identical quantum particles have not been misguided
from the very start. 
\end{abstract}

\section{Introduction}
In classical physics, particles are the example \emph{par
excellence} of distinguishable individuals. No two classical
particles can be in exactly the same physical state: in Newtonian
spacetime different particles will at least occupy different
spatial positions at any moment, because of their impenetrability.
They will therefore obey Leibniz's Principle of the Identity of
Indiscernibles, which says that different individuals cannot share
all their physical properties. Moreover, classical particles
possess \emph{genidentity}, i.e.\ identity over time. That is,
given two particle configurations at different instants, it is an
objective physical fact for each particle at the later instant
with which particle in the earlier configuration it corresponds.
This is because classical particles follow definite trajectories
that make it possible to follow them over time. Classical
particles can thus always be distinguished and be given individual
names, or numbers: particle $1$, particle $2$, etc. These particle
numbers are correlated with different, identifying physical
characteristics.

In quantum theory the status of individual objects is a
notoriously more complicated subject. The standard quantum
mechanical treatment of particles starts simply enough, with the
uncontroversial case of one particle described by a state in a
single Hilbert space. In the case of two or more particles the
tensor product of such individual Hilbert spaces is formed,
$\mathcal{H}_1 \bigotimes \mathcal{H}_2 \bigotimes \mathcal{H}_3
\bigotimes ...$. The natural interpretation, especially with the
classical case in mind, is that in such formulas $\mathcal{H}_i$
is the state space of particle $i$. In other words, it seems
natural to interpret the indices as not only referring to the
individual factor spaces in the total tensor product Hilbert
space, but also to individual \emph{particles}.

Complications arise, however, for particles of the same kind
(so-called ``identical quantum particles''). In this case it is a
basic principle of quantum theory that only completely symmetrical
or completely antisymmetrical states are allowed in the tensor
product space $\mathcal{H}_1 \bigotimes \mathcal{H}_2 \bigotimes
\mathcal{H}_3 \bigotimes ...\bigotimes \mathcal{H}_n$. In such
fully (anti)sym\-metrical states the restriction of the state to a
single factor space (i.e., the density operator obtained by
``partial tracing'' over the variables of the other factor spaces)
is the same for all factor spaces. In other words, all
one-particle states, defined in the individual Hilbert spaces
$\mathcal{H}_i$, are all equal. If the indices $i$ are regarded as
particle indices, this means that the several particles cannot be
distinguished on the basis of their state-dependent properties
(like position, momentum, etc.)\footnote{Here we follow standard
interpretational ideas, according to which the states we just
mentioned provide \textit{complete} physical descriptions. If, on
the other hand, it is assumed that a finer description is
possible, and that the quantum state only provides statistical
information about the actual properties of physical systems---like
in Bohm's theory or in modal interpretations---there may very well
exist individuating physical characteristics. The whole issue is
therefore interpretation dependent; our discussion here stays
within the same standard interpretational framework adopted by
most recent discussions about the individuality of identical
particles.}. Since the state-independent properties (charge, rest
mass, etc.) are by definition equal for ``identical particles'',
this leads to the conclusion that \textit{all} particles of the
same kind possess exactly the same physical properties. Their
individuality can therefore not be based on individuating physical
properties, and Leibniz's Principle is therefore apparently
violated. This strange situation is the origin of a very extensive
literature about the nature of identical quantum particles. At the
present moment the main theses discussed in this literature are
that the individuality of identical particles must itself be a
fundamental property (``haecceity'', ``fundamental thisness'') in
the case of bosons, since there are no ordinary physical
differences between them at all; and that in the case of fermions
a form of Leibnizean discernibility can be salvaged (``weak
discernibility'') because in this case there are irreflexive
relations between the particles (see
\cite{french,vanfraassen,dieks0} for general discussion,
\cite{muller1,muller2,saundersleibn,saundersobj,teller} for
elaboration of the just-mentioned positions, \cite{dieks,dieks1}
for criticism).

It is important to note that the symmetrization postulates, which
are responsible for the equality of all one-particle states, are
basic postulates of quantum mechanics that apply to the collection
of \emph{all} particles of the same kind. This means, for example,
that \emph{all} electrons in the universe are in exactly the same
state, whatever the differences are between the physical
conditions at different positions in the universe. In fact, it
does not make sense to distinguish between electrons here and
electrons elsewhere, for instance in another solar system: all
electrons in the universe are ``partly here, partly there, a bit
everywhere'' (see below for more on this). It is not relevant for
this universal applicability of the symmetry postulates what kinds
of interactions and situations are considered; in particular,
whatever circumstances are important for the transition to the
classical limit, these do not affect the applicability of the
symmetrization postulates. This implies that even in the classical
limit the different particle indices $i$ are all associated with
exactly the same state. In other words, it seems that even
classical particles must be completely indistinguishable!

This result is obviously highly problematic---in fact, we began
our whole argument by pointing out that classical particles are
distinguishable objects \emph{par excellence}. So something must
have gone wrong at some point in the above reasoning. In the
remainder of this article we shall analyze the source of this
paradox, and investigate the implications of our analysis for the
particle concept in quantum mechanics.

\section{The states of identical particles}\label{EPR}

Consider the concrete case of a system consisting of two
electrons\footnote{As just explained, we should in principle
always consider the fully entangled state of \emph{all} electrons
in the universe and, in view of the equality of all partial
traces, considering a two-electrons \textit{sub}system with
specific properties does not really makes sense then. So at the
moment it is best to think of a universe in which there exist only
two electrons. In sections \ref{classpart1} and \ref{classpart2}
we shall work out a particle concept for which it does make sense
to consider specific subsystems.}. Electrons are fermions and
therefore have an antisymmetrical state, typically looking like
\begin{equation}\label{twoelectrons}
| \Psi \rangle = \frac{1}{\sqrt{2}}(| \phi_1 \rangle | \psi_2
\rangle - |\psi_1 \rangle | \phi_2 \rangle ).
\end{equation}
Here, the subscripts $1$ and $2$ indicate whether the indexed
state is defined in $\mathcal{H}_1$ or in $\mathcal{H}_2$,
respectively. Taking partial traces in $|\Psi\rangle$, we find
that both the state restricted to $\mathcal{H}_1$ and the state
restricted to $\mathcal{H}_2$ has the form $W = \frac{1}{2} ( |
\phi \rangle \langle \phi | + | \psi \rangle \langle \psi | )$. If
we think of $\mathcal{H}_1$ and $\mathcal{H}_2$ as the state space
of particle $1$ and particle $2$, respectively, we can thus
conclude that both particles are in exactly the same state:
figuratively speaking, they are both half in $| \phi \rangle$ and
half in $| \psi \rangle$.

This means that it would not be correct to say that in state
(\ref{twoelectrons}) there is one particle in $| \phi \rangle$ and
one in $| \psi \rangle$. In fact, a state with particle $1$ in $|
\phi \rangle$ and particle $2$ in $|\psi \rangle$ necessarily
would have to possess the product form $| \phi_1 \rangle \otimes |
\psi_2 \rangle$ \cite[sect.\ VI.2]{vonneumann}, which not only
conflicts with the symmetrization postulate but is also
empirically different from (\ref{twoelectrons}). The expectation
value of an observable of the two-electron system (a symmetrical
hermitean operator) $A$ in state (\ref{twoelectrons}), $\langle
\Psi |\, A \, | \Psi \rangle$, differs from its expectation value
in a product state by the presence of an interference term
$\langle \phi_1 \otimes \psi_2|\, A \, | \psi_1 \otimes
\phi_2\rangle$. It may happen, of course, that this cross term
vanishes for particular choices of $A$, and in this case using the
product state does not bring us into conflict with empirical
results. But then there always are other observables for which the
cross terms do not vanish, and empirical evidence confirms the
existence of these terms. This means that the suspension of the
symmetrization postulates that sometimes occurs in the physics
literature, for instance when spatially isolated systems are
subjected to position measurements, is only pragmatically
justified. This manoeuvre simplifies the calculations, but has no
fundamental status. The fully symmetrized entangled state has
general applicability and validity, and is therefore the only one
to be used in a general conceptual analysis.

The conclusion is thus unavoidable that different ``particle
indices'' $i$, $j$ cannot be associated with any measurable
physical differences. Among philosophers of physics this is an
acknowledged fact that has given rise to the hotly debated
question of what then \emph{is} able to ground the individuality
of these particles.

But it should be noted that in those parts of the foundational
literature that do not focus on identity issues, and in the actual
practice of physics, the use of the particle concept is not
unequivocal. It is true that particles are sometimes associated
with the indices of our above discussion; but one also encounters
another, very different use of the particle concept (cf.\
\cite{lubberdinka,lubberdinkb}, where a distinction is introduced
between ``h particles'' and ``q particles''). This alternative
approach is significant for our later analysis, and we therefore
want to illustrate it by an example, namely the notorious
Einstein-Podolsky-Rosen case.

In its modern version, the EPR experiment refers to two electrons
``at a distance from each other'', on which spin measurements are
performed. The two-particle spin state on which attention usually
focusses is the singlet state, but the full state obviously also
contains a spatial part. The correctly symmetrized total state is
\begin{equation}\label{EPReq}
| \Phi \rangle = \frac{1}{\sqrt{2}}(| \phi_1 \rangle | \psi_2
\rangle + |\psi_1 \rangle | \phi_2 \rangle ) (|\!\uparrow_1
\rangle |\downarrow_2 \rangle - |\!\downarrow_1 \rangle
|\uparrow_2 \rangle ),
\end{equation}
where $|\!\!\uparrow \rangle $ and $|\!\!\downarrow \rangle $
stand for spin eigenstates with spin directed upwards and
downwards in a particular direction, respectively, and where $(|
\phi \rangle$ and $| \psi \rangle$ now refer to states that are
localized ``on the left hand side'' and ``the right hand side'',
respectively. In the language of wave mechanics, $| \phi \rangle$
and $|\psi \rangle$ represent localized wave packets at a
macroscopic distance from each other.

In the literature on EPR the state is very often given in a
different form, namely
\begin{equation}\label{EPRvar}
| \Phi^{\prime} \rangle = \frac{1}{\sqrt{2}} |\phi_1 \rangle |
\psi_2 \rangle (|\!\uparrow_1 \rangle |\downarrow_2 \rangle -
|\!\downarrow_1 \rangle |\uparrow_2 \rangle ),
\end{equation}
in which the spatial part is a simple product state. Clearly, this
state does not obey the anti-symmetrization postulate, and from a
fundamental point of view it therefore cannot be right. It is true
that as long as we only consider observables that commute with
position, we shall not find any predictions that differ from those
found from Eq.\ (\ref{EPReq}), and this yields a pragmatic
justification for using (\ref{EPRvar}). However, the really
important advantage of using Eq.\ (\ref{EPRvar}) instead of the
correct state is that this form of the state lends itself to an
easy interpretation: we have one particle at the left hand side,
and one on the right hand side. This, of course, fits in with the
standard way of speaking about EPR. According to the usual
discussions there is a left-side particle $L$ and a right-side
particle $R$, and we are interested in the results of spin
measurements on these two individual particles. Note that in this
common way of dealing with the situation the particles are
considered as individuals that differ from each other in their
physical properties, namely their locations.

But if we use the correct form (\ref{EPReq}), and associate our
particles, in accordance with the official doctrine, with the
``particle indices'' $1$ and $2$, we have to conclude that there
\emph{is} no left and no right particle: the states of both $1$
and $2$ are ``evenly distributed'' between left and right. This
means that the way the EPR case is usually understood, as being
about a particle $L$ and a particle $R$, is at variance with the
official doctrine regarding the concept of particles in quantum
mechanics.

Of course, those who think in terms of individual localized
particles in this case (i.e., in practice almost everyone)
generally \emph{know} that the state in principle has the form
(\ref{EPReq}); but this does not induce them to abandon the idea
of an individual $L$ and an individual $R$ particle. This points
into the direction of the existence of an alternative way of
handling the particle concept, one that does \emph{not} relate
particles to the indices in the tensor product formalism.
Apparently, such an alternative conception is already present in
the practice of physics---at least on an intuitive level. As we
shall see, if worked out this other way of interpreting the
particle concept, rather than the official doctrine that
\emph{indices} represent particles, provides a natural bridge to
particles as they occur in classical physics.

\section{Classical particles in quantum mechanics}\label{classpart1}

Classical particles are characterized by their unique spatial
positions and trajectories. It is often said that both these
features are excluded in quantum mechanics. For example, in many
textbooks the statement can be found that in quantum mechanics it
cannot be an objective fact which particle at a later instant is
identical with which earlier particle, because of the absence of
trajectories. According to this argument the concept of
genidentity does not apply to quantum particles. However, if this
absence of particle localization and particle trajectories were a
matter of principle, the resulting situation would be very
puzzling. Surely, the classical particle picture must be expected
to emerge from quantum mechanics in some limiting case, and one
must therefore assume that the typical classical particle features
can be mimicked in quantum mechanics. In fact, that this is indeed
the case is well known, in spite of the declarations to the
contrary that we have just mentioned. One key result in this
connection is Ehrenfest's theorem about the dynamics of
expectation values of observables.

In the case of a Hamiltonian $H = p^2 /2m + V(r)$, with $p$ the
momentum, $m$ the particle mass and $V(r)$ a potential field, we
can introduce a force field $F(r) = -\nabla V(r)$, in terms of
which Ehrenfest's theorem takes the form
\begin{equation}\label{Ehrenfest}
\langle F(r) \rangle = m \frac{d^2}{dt^2} \langle r \rangle .
\end{equation}
For certain specific potentials (free motion, i.e.\ F=0, or if V
is a quadratic function of $r$) we find that $\langle F(r)
\rangle$ equals $ F(\langle r \rangle)$, so that in these cases
the mean value of $r$ exactly satisfies the classical law of
motion $F(\langle r \rangle) = m \frac{d^2}{dt^2} \langle r
\rangle$. In general this is not so. But if the wave function is
localized in a sufficiently small region of space, so that the
variation of the force field within that region is small, we can
replace Eq.\ (\ref{Ehrenfest}) by the classical equation in a good
approximation (which becomes better when the state becomes more
localized). From this it follows that well-localized
single-particle quantum states (localized in the sense that their
associated wave packets are very narrow) follow classical
trajectories to a very good approximation.

Classical trajectories therefore do exist in quantum mechanics:
they are realized by (very) small wave packets. In the case of a
Hamiltonian that is quadratic in position---the harmonic
oscillator being the prime example---such small wave packets
remain small over time: their widths merely oscillate. This case
therefore furnishes an example of a quantum system that almost
perfectly mimics the behavior of a classical particle.

If the potential does not have this special form there will in
general be dispersion, so wave packets will spread out. Classical
motion will then only be a good description of the behavior of the
average position of the wave packet during a finite time, during
which the approximation $\langle F(r) \rangle = F(\langle r
\rangle)$ remains valid. Moreover, even if the center of the wave
packet stays on a classical trajectory, the analogy with a
classical particle path will get partially lost if the packet
becomes too extended. Free motion is the obvious example: although
in this case the average position of a moving wave packet will
always be exactly on the classical trajectory, the width of the
packet will increase in an approximately linear way, according to
$\triangle r = \{(\triangle r_0)^2 + (\triangle p_0 t/m)^2
\}^{1/2}$ (with $t$ representing time). When the size of the
packet has become substantial, the results of consecutive position
measurements will no longer need to lie on a classical path, not
even approximately. A classical particle picture therefore does
not apply in this situation. Consequently, we need a mechanism to
keep wave packets narrow in order to maintain classical
particle-like structures in quantum mechanics over longer
stretches of time.

Such considerations are standard in studies on the classical limit
of quantum mechanics, and there is growing agreement that the
essential element in explaining how classical mechanics emerges
from quantum mechanics is the process of decoherence. The key
ideas are that physical systems are usually not isolated but
interact with an environment; and that in many circumstances the
interaction is such that the environment effectively performs
(approximate) position measurements on the systems in question.
The effect of this interaction with the environment is the
destruction of coherence between parts of the wavefunction
centered around different positions: these parts become correlated
with mutually orthogonal environment states. Consequently,
spatially extended wave functions are transformed into mixtures of
spatially very narrow states. Model calculations indicate that
these narrow wave functions obey the quantum mechanical evolution
equation governed by the system's own Hamiltonian (leading, among
other things, to the validity of Ehrenfest's theorem commented
upon above) plus two terms representing the interaction with the
environment (see \cite{zurek}, especially equations $17$ and $24$
therein). The first of these terms is a damping term, representing
friction with the environment; the second term, more important for
our purposes, represents the decoherence process and keeps on
minimizing the dimensions of the wave packet.

As a result, the classical limit of quantum mechanics is
characterized by the emergence of classical particle trajectories
that are followed by narrow wave packets. These narrow, localized
wave packets become the particles we are familiar with in
classical physics. Collections of such localized wave packets
represent the particle subsystems we commonly refer to in the
classical context (compare footnote $2$ in section \ref{EPR}).

\section{The particle concept in quantum mechanics}\label{classpart2}

The finer details of the decoherence mechanism, and the work that
remains to be done to fully understand them, need not detain us
here. The important thing is that there is a consensus that
classical particles emerge from quantum mechanics as narrow wave
packets that in very good approximation follow classical particle
paths. This is the conceptual background of what we have signalled
before: in the practice of physics the particle concept is very
often not linked to the indices in the formalism, but rather to
distinct localized states. The EPR experiment, where the localized
states on the left and right wing of the experiment are associated
with an individual $L$ and $R$ particle, respectively, is but one
example of this. The way we speak about experiments (the positrons
in the CERN experiment, etc.) or about the objects surrounding us
(the quantum particles making up this table) are other examples.

Thinking about quantum particles in this way is eminently
reasonable. The origin of the concept ``particle'' comes from
classical physics, and in this classical context we know exactly
what we are talking about when we use the term. In fact, our
language is permeated by the concepts of localized objects and
particles. Given this background knowledge, it seems a matter of
course to reserve the same term in quantum mechanics for things
that share core characteristics with classical particles and that
at least become recognizable as classical particles in the
classical limit. This bill is fitted by localized states, but not
by the states associated with ``particle indices''.

To emphasize the difference between the two rival quantum particle
concepts once again, we want to take another look at a symmetrized
many-particle state. Suppose that in the state
\begin{equation}\label{twoparticles}
\frac{1}{\sqrt{2}}(| \phi_1 \rangle | \psi_2 \rangle + |\psi_1
\rangle | \phi_2 \rangle ) \end{equation} the one-particle states
$| \phi \rangle$ and $| \psi \rangle$ are localized. It could be
that they have always been this way, and that the dynamics
preserves the localization (as in the harmonic oscillator case),
but the more typical situation is that decoherence processes are
responsible for the appearance of this localized two-particle
state in a incoherent mixture of similar states. Now, according to
the way of handling the particle concept that we have just
explained, the state (\ref{twoparticles}) represents two
individual and distinguishable particles, one at the position
where $| \phi \rangle$ is localized and the other at the position
defined by $| \psi \rangle$. By contrast, if we hold fast to the
idea that particles are represented by the indices occurring in
the formalism, we arrive at the conclusion that
(\ref{twoparticles}) represents a situation in which there are two
\emph{indistinguishable} ``particles'', both in the state
$\frac{1}{2} ( | \phi \rangle \langle \phi | + | \psi \rangle
\langle \psi | )$. As we have stressed before, this
indistinguishability survives the classical limit: since all
factor Hilbert spaces and the states defined in them occur
completely symmetrically in the total state, all interactions will
affect the states in the factor spaces in exactly the same way. So
all indices will remain associated with the same density operator,
evenly distributed over the pure one-particle states. The
``index-particles'' therefore do not become classical particles in
this limit: they refuse to become localized. This seems a
\emph{reductio} of the idea that the Hilbert space indices can be
taken to stand for particles.

\section{Emergence of particles in quantum mechanics}

Our proposal is therefore to think of particles in quantum
mechanics as represented by localized wave packets. That is to
say, if we encounter a state $|\Psi\rangle$ defined in
$\mathcal{H}_1 \bigotimes \mathcal{H}_2 \bigotimes \mathcal{H}_3
\bigotimes ...\bigotimes \mathcal{H}_n$, and wish to investigate
whether it can be interpreted in terms of particles, we have to
ask ourselves whether it can be written as a symmetrized product
of localized one-particle states. A worry that might arise here is
whether such a decomposition of $|\Psi\rangle$, if it exists, is
unique. If more than one particle-like representations of
$|\Psi\rangle$ could be found, the uniqueness of the classical
limit and the meaningfulness itself of our particle concept would
be endangered. At first sight this worry seems certainly serious,
because the symmetrization postulates require that the
coefficients appearing in front of the product terms in the
symmetrized state $|\Psi\rangle$ are all equal. For example, in
state (\ref{twoparticles}) both terms are prefixed by
$\frac{1}{\sqrt{2}}$, which means that we are dealing with a
degenerate Schmidt (bi-orthogonal) decomposition. In such a case
there are infinitely many other Schmidt decompositions: each
rotation in the subspaces of $\mathcal{H}_1 $ and $ \mathcal{H}_2$
spanned by $| \phi \rangle$ and $| \psi \rangle$ leads to a new
pair of vectors $| \phi^{\prime} \rangle, | \psi^{\prime} \rangle$
in terms of which the bi-orthogonal form (\ref{twoparticles}) can
be written down too. However, and this is crucial, these
alternative decompositions will not be in terms of
\emph{localized} wave packets. Indeed, rotations in Hilbert space
are implemented by unitary transformations that transform the
original vectors into linear combinations of them; if the original
states are localized in connected regions of space the transformed
stated, being superpositions of the original ones, are obviously
not thus localized. It follows that if a decomposition of state
$|\Psi\rangle$ is possible in form (\ref{twoparticles}) with
localized states $| \phi \rangle$ and $| \psi \rangle$, this
decomposition is unique. This argument generalizes immediately to
the case of an arbitrary number of particles.

The demand that a state represents \emph{particles}, in the sense
we have defined here, is therefore much stronger than that the
state can be written in the form (\ref{twoparticles}) with
mutually orthogonal $| \phi \rangle$ and $| \psi \rangle$. The
latter is always possible, for any state in a Hilbert space
$\mathcal{H}_1 \bigotimes \mathcal{H}_2$ (because of Schmidt's
theorem). It is the added localizability condition that makes the
question of whether there exists a decomposition of the required
form non-trivial, and makes the decomposition unique \emph{if} it
exists.

In most cases states will \emph{not} allow a particle
interpretation; think, for example, of a state of the form
(\ref{twoparticles}) with two overlapping wave packets $| \phi
\rangle$ and $| \psi \rangle$ (each defined in a connected region
of space). The bi-orthogonal decomposition that we need, in terms
of localized states that are non-overlapping (and therefore
mutually orthogonal) clearly does not exist: there of course does
exist a bi-orthogonal decomposition, but the states occurring in
it will be linear combinations of $| \phi \rangle$ and $| \psi
\rangle$ and will therefore overlap spatially. An arbitrarily
chosen quantum state will therefore not describe particles. We
need special circumstances to make the particle concept
applicable. In this sense, the classical limit with its
decoherence processes makes classical particles really
\emph{emerge} from the substrate of the quantum world.

It should be added that the circumstances that are responsible for
the emergence of classical particles also justify the use of the
statistics that we expect for the case of several independent
individuals. The symmetrization postulates require that
``many-particle states'' (in the sense of general states in a
Hilbert space that is the tensor product of more than one factor
spaces) are entangled, and in general this leads to the existence
of correlations in measurement results, even if there is no
question of past or present mutual interactions. From our
perspective, this remarkable ``quantum statistics'' (either
Fermi-Dirac or Bose-Einstein) points into the direction of a
failure of the individual particle concept in the general quantum
situation. The ``particle alternative'' is to see the existence of
these correlations as a sign that quantum particle states are
subject to peculiar initial conditions (see
\cite{french,vanfraassen,dieks0} for discussion), or that quantum
particles exert ``exchange forces'' on each other (repulsion
between fermions and attraction between bosons, see \cite{mullin}
for a critical discussion of this concept). In our approach this
complication does not arise, since we reject the idea of particles
in the general situation in which we do not have localized
systems. In the case of spatially non-overlapping wave packets, in
which the particles concept does become applicable, both
Fermi-Dirac and Bose-Einstein statistics reduce to the usual
Boltzmannian statistics, as is well known.

\section{Classical particles and indices}

In order to obtain a clearer view on the relation between
particles as we have defined them here, via localized wave
packets, and the usual particle concept that is linked up with the
indices in the formalism, it is helpful to compare with an
analogous situation in classical mechanics
\cite{lubberdinka,lubberdinkb}. As it turns out, it is possible to
define classical ``indistinguishable particles'' that are
analogous to the indistinguishable ``index-particles'' in quantum
mechanics.

We do not ordinarily use symmetry postulates in classical
mechanics, but in the case of particles of the same kind we could
introduce a symmetrization procedure without changing the
empirical content of the theory. In the usual formalism the state
of a system consisting of $n$ particles of the same kind is
represented by one point in phase space, with the first coordinate
standing for the position of particle $1$, the second for the
position of particle $2$, the $n+1$-th coordinate representing the
momentum of particle $1$, etc. Obviously, it would not make any
empirical difference if we would call particle $1$ particle $2$,
etc. The only thing that is important is that there is one
particle in state $(x_1,p_1)$, one in $(x_2.p_2)$, etc.; the
states individuate the particles and it is irrelevant how we
\emph{call} them. Permutation of the \emph{names} of the particles
will not lead to any physical differences. With this in mind,
consider all permutations of the particle numbers, in which these
are distributed differently over the one-particle states. This
will generate $n!$ phase points, in which the individual
one-particle states are numbered differently, corresponding to the
number (name) of the particle to which they pertain. These $n!$
states are all empirically indistinguishable from each other and
from the original state, the only difference between them being
the way the one-particle states are indexed, i.e.\ on which phase
space axes each particular one-particle state $(x,p)$ is
represented. Now, instead of the usual mechanical state, given by
our single original phase point, we might introduce a
\emph{symmetrized} state represented by the complete collection of
these $n!$ points. This new state is symmetrical because it is
invariant under permutations of the indices. All the usual
formulas from classical mechanics can be reformulated to
accommodate this new state definition: the idea simply is to do
the usual calculations for each point separately. For the case of
dynamical evolution this will lead to a new $n!$-points state,
again symmetrical and with all points empirically equivalent; and
in general, the calculations will lead to $n!$ results that only
differ from each other in their assignments of indices. The final
result can then be taken as the collection of these $n!$ partial
results.

Evidently, the sole purpose of this manoeuver is to make it
manifest that nothing physical depends on an arbitrary numbering
of the particles. The particles are physically characterized by
the individual states $(x,p)$, not by the indices. But now suppose
that, in spite of this symmetrization, we are caught in the idea
that each \emph{index} has to correspond to a specific particle,
and that we are going to inquire about the state of particle $i$.
In our symmetrized scheme, the natural answer consists in the
collection of $(x,p)$ states that bear the index $i$, given the
$n!$ phase points that make up the many-particle state. In this
way \emph{all} one-particle states are attributed to \emph{each}
index value. The conclusion would then be that the particles are
all in exactly the same state and therefore indistinguishable.

It goes without saying that the latter conclusion is in conflict
with the way the particle concept is actually used in classical
mechanics. In classical mechanics particles are as distinguishable
as their states $(x,p)$ are; and from this point of view the above
argument is simply a confused misinterpretation of the indices.
The indices were only formal expedients, but the argument took
them to denote individual physical objects. The resulting
indistinguishability paradox is dispelled once we realize what
role the indices really play.

This mistaken piece of classical arguing is, however, analogous to
the standard reasoning we find in quantum mechanics: instead of
looking for individuating physical particle characteristics that
might make it possible to speak of individual, distinguishable
entities, one holds fast to the \textit{a priori} idea that the
Hilbert space indices should play this role. The symmetry of the
formalism should give one pause: instead of indicating that all
particles are in the same state, this symmetry signals that the
indices do \emph{not} have the role of particle names.

\section{Weak discernibility}

In the foregoing sections we have focused on the classical limit
of quantum mechanics, with the aim of showing that the indices in
the tensor product formalism do not become classical particle
names in this limit. This leaves it open, however, that the
indices refer to individual objects of a different kind, genuine
\emph{quantum} particles, say, that remain distinct from the
\emph{classical} particles that emerge in the classical limit.
There is indeed a growing literature in which it is claimed that
the indices in the tensor product formalism do refer to individual
physical entities, distinct from each other by virtue of their
\emph{physical characteristics}. This is clearly a claim that has
to be investigated. Associating each index with its own
\emph{haecceity} in order to guarantee that it corresponds to an
individual object is a move that needs not be taken very
seriously, as it boils down to attributing individuality by
\emph{fiat}; but if it is true that there are physical features
that individuate the indices, the conclusion that they are
denoting individual quantum objects becomes harder to resist.

We have already seen that all indices are associated with exactly
the same reduced state; this seems to make the existence of
individuating physical properties impossible from the outset.
However, there is a way out on which the literature we just
alluded to is based. The core idea is in the observation that even
within the scope of classical physics situations are thinkable in
which entities are in identical states but are nevertheless
distinct individuals, namely situations with complete symmetry. A
famous example was introduced by Max Black \cite{black}: consider
two spheres of exactly the same form and material constitution,
alone in a relational space (in order to exclude absolute position
as a distinguishing property), at a fixed distance from each
other. This is a situation that seems certainly thinkable without
getting into contradictions. But it is also a situation in which
no physical features are able to distinguish between the two
spheres, in spite of the fact that there are obviously two of
them. The spheres thus seem to defy Leibniz's Principle, and
appear to possess an identity that cannot be grounded in physical
differences.

However, there is a way to save a form of Leibniz's Principle in
such symmetrical classical configurations. As pointed out by
Saunders \cite{saundersleibn,saundersobj}, who takes his clue from
Quine \cite{quine}, \textit{irreflexive} relations are
instantiated here: relations entities cannot bear to themselves.
In the case of the spheres, each sphere has a non-vanishing
distance to one other sphere; and an object cannot possess such a
distance to itself. This irreflexivity is the key to proving that
(a generalized version of) Leibniz's Principle is satisfied after
all. If an entity stands in a relation that it cannot have to
itself, there must be at least two entities. It is not difficult
to formalize this argument and to prove that the existence of
irreflexive physical relations grounds the multiplicity of
objects, without recourse to haecceities. Because of the sameness
of all individual states it still is impossible to give
\emph{names} based on physical characteristics in such cases; for
example, we cannot give a description of one of our spheres that
would not apply equally to the other one. The objects are
therefore not distinguishable in the usual sense; but still we can
prove that there are more than one of them. For this reason the
term ``weak discernibility'' has been introduced to capture how
objects differ from each other in such cases.

The idea now is that in quantum mechanics the situation is
analogous. That is, although the states associated with different
indices are identical, irreflexive relations exist between the
indices that make them weakly discernible. In the case of fermions
the total state is antisymmetric, like in Eq.\
(\ref{twoelectrons}), and here the irreflexive relation takes the
form of ``being associated with different one-particle states''.
Indeed, as can be verified in Eq.\ (\ref{twoelectrons}), in the
antisymmetric case each term of the superposition contains indices
that indicate different vectors in Hilbert space. From this
Saunders \cite{saundersleibn,saundersobj,muller1} concludes that
fermions are physical \emph{individuals}, albeit only weakly
discernible ones. Muller and Seevinck extend this argument to
bosons \cite{muller2}. They observe that quite generally there
exist irreflexive relations between the indices: operators that
belong to different Hilbert spaces, indexed by different indices,
always commute, whereas this is not the case for operators
belonging to the same Hilbert space. In particular, momentum and
position operators with different indices commute, whereas they do
not if their indices are the same. So even bosons appear to be
weakly discernible individuals.

It should be noted, however, that these arguments hinge on a
silent premiss, namely that the indices not only play a
mathematical role but also possess \emph{physical} significance.
As mathematical demonstrations, demonstrating the individuality of
the different factor spaces, they are unproblematic; but we need
an additional justification for thinking that the indices also
correlate to something physical. Of course, it is simple enough to
find irreflexive relations between \emph{numbers}, for instance
the relation of inequality. It is also easy to couch such
relations in physical language, for instance by speaking about
\emph{observables} that belong to the same or different Hilbert
spaces. But that will not suffice, we need a positive indication
of physical relevance---certainly not all mathematical quantities
occurring within a physical theory refer to things existing in the
physical world. In our case, whether the indices and the
quantities labelled by them possess physical significance is
precisely the issue under discussion, and it would therefore be
question begging to assume this significance. We need an argument
to make the indices physically respectable. To get a clue about
possible criteria here, let us first have a look at classical
physics.

In situations in classical physics without particular symmetries,
physical relations can be used to distinguish and name the things
that are related. For example, in an arbitrary configuration of
more than two classical particles the distances with respect to
the other particles will unambiguously characterize each
individual particle. Changing the configuration so that it becomes
more symmetrical (but not yet completely symmetrical) will change
the values of the distances, but not the number of individual
objects. Distance relations thus are the kind of relations that
connect actual physical objects. This possibility of
distinguishing and naming actual objects in asymmetrical
situations provides us with a justification for considering
distance relations as physically meaningful, in the sense that if
these relations apply, they apply to physical objects. The
completely symmetrical situation is a degenerate situation, a
limiting case, in which naming via distances admittedly becomes
impossible but in which the distance relations are still
sufficient to establish \emph{weak} discernibility and are able to
fix the \emph{number} of objects.

Indeed, why are we so sure intuitively that there are two Blackean
spheres? This is because our mind's eye sees these spheres at
different distances or in different directions before us; in
thought we break the symmetry, which makes it possible to
distinguish the entities and name them (the left and right sphere,
for example). The symmetrical configuration is thus thought of as
a limiting case of the more familiar asymmetrical situation.

Now compare the quantum case. Can a similar story be told here, to
make it acceptable that the indices are potential particle labels?
Unfortunately, this attempt is immediately thwarted by the
symmetrization postulates. It is a \emph{fundamental principle} of
quantum theory that the indices can \emph{never} appear in
configurations that are not symmetrical. In classical physics
perfect symmetry of particle configurations, if it occurs at all,
is something contingent; but in quantum mechanics it is a law-like
feature that all indices must always occur, in any expression and
in whatever situation, in a fully symmetrical way. It is even
useless to introduce an external standard: if in thought we inject
ourselves into the world of electrons, quantum theory requires
that all relations between us and the electrons remain completely
symmetrical in the indices. This is very much different from the
case of Black's spheres. In quantum mechanics it is a matter of
\emph{principle } that we can \emph{never} associate different
physical characteristics with different indices in the formalism.
We therefore lack evidence that the indices may refer to distinct
physical entities at all (see for more on this
\cite{dieks,dieks1}). The irreflexive relations in which the
indices stand can therefore not be assumed to connect physical
entities.

But what about our actual experience, telling us that in many
experiments we do encounter individual electrons and other
particles? This we have discussed in the previous sections: such
experiments, to the extent that they provide convincing evidence
about the presence of particles, pertain to classical limiting
situations. As we have seen, the particles that emerge in those
situations do \emph{not} correspond to the indices in the quantum
formalism.

\section{Conclusion}

We conclude that the indices in the quantum mechanical formalism
of ``identical particles'' refer to the individual factor spaces
from which the total Hilbert space in the formalism is
constructed---they are merely \textit{mathematical} quantities.

In order to support this conclusion we have first argued that,
within a standard no-hidden-variables interpretational context,
the classical limit does not associate Hilbert space indices with
particles as we know them from pre-quantum physics. However,
well-localized wave packets do take on this role: they do
represent classical particles in the limiting situation. The
appearance of particles in quantum mechanics is therefore a case
of emergence. Only if specific physical conditions are satisfied,
resulting in the presence of localized wave packets (decoherence
processes are usually essential here) does the concept of a
particle in the ordinary sense become applicable to the world
described by quantum mechanics. As just said, these emerging
particles are not linked to the indices occurring in the
formalism.

Second, we have argued more generally that in the standard
interpretation there is no indication that the indices in the
formalism denote distinct physical entities at all. Rather, the
symmetrization postulates have the effect of eliminating any
potential physical label-like role of the indices. The analogy
between quantum mechanical systems of ``identical particles'' and
classical collections of symmetrically positioned weakly
discernible objects is only superficial. There is no sign within
standard quantum mechanics that ``identical particles'', denoted
by indices, are physical objects at all.

This conclusion raises the question of whether the discussions in
the philosophy of physics about the nature of the individuality of
identical quantum particles have not been misguided. It makes
little sense to wonder whether Leibniz's Principle is satisfied by
identical quantum particles, or whether they possess haecceities,
if the existence itself of these particles has not been
established.

\end{document}